\documentclass{article}

\usepackage{arxiv}

\usepackage{verbatim}
\usepackage{epsfig}
\usepackage{amsmath}
\usepackage{amsthm}
\usepackage{amssymb}
\usepackage{psfrag}
\usepackage[T1]{fontenc}
\usepackage[ruled]{algorithm2e}
\usepackage{textcomp}

\usepackage[usenames,dvipsnames]{pstricks}
\usepackage{pst-grad} 
\usepackage{pst-plot} 
\usepackage[ruled]{algorithm2e}
\usepackage{etoolbox}
\usepackage{subfig}
\makeatletter
\patchcmd{\@makecaption}
  {\scshape}
  {}
  {}
  {}
\makeatother

\newtheorem{theorem}{Theorem}
\newtheorem{lemma}{Lemma}

\newtheorem{definition}{Definition}
\newtheorem{assump}{Assumption}

\newtheorem{proposition}{Proposition}
\newtheorem{corollary}{Corollary}[theorem]
\newtheorem{remark}{Remark}[theorem]

\newcommand{\G}{\mathcal{G}}
\newcommand{\V}{\mathcal{V}}
\newcommand{\E}{\mathcal{E}}
\newcommand{\Pd}{\mathcal{P}}
\newcommand{\bigO}{\mathcal{O}}

\definecolor{color1}{rgb}{0,0,0}

\title{Distributed finite-time termination for consensus algorithm in switching topologies}

\author{
  Govind Saraswat, Vivek Khatana, Sourav Patel, Murti V. Salapaka \\
  Department of Electrical and Computer Engineering\\
 University of Minnesota\\
 Minneapolis, USA\\
  \texttt{\{saras006,khata010,patel292,murtis\}@umn.edu} \\
}

\begin{document}
\maketitle
\date{}

\begin{abstract}
In this article, we present a finite time stopping criterion for consensus algorithms in networks with dynamic communication topology. 
Recent results provide asymptotic convergence to the consensus algorithm. However, the asymptotic convergence of these algorithms pose a challenge in the practical settings where the response from agents is required in finite time. 
To this end, we propose a Maximum-Minimum protocol which propagates the global maximum and minimum values of agent states (while running consensus algorithm) in the network. We establish that global maximum and minimum values are strictly monotonic even for a dynamic topology and can be utilized to distributively ascertain the closeness to convergence in finite time. 
We show that each node can have access to the global maximum and minimum by running the proposed Maximum-Minimum protocol and use it as a finite time stopping criterion for the otherwise asymptotic consensus algorithm. The practical utility of the algorithm is illustrated through experiments where each agent is instantiated by a NodeJS \textit{socket.io} server.
\end{abstract}

\keywords{Distributed Consensus\and Switching topology\and Multi-agent systems\and Network-based computing systems}

\section{Introduction}

Availability of large number of low-cost sensors and development of suitable network protocols has led to the development of modern-day multi-agent systems. In many practical domains, these multi-agent systems are often designed to attain coordinated objectives such as movement coordination among a group of mobile autonomous vehicles for traffic optimization \cite{jadbabaie2003coordination}, task allocation among agents (nodes) such as unmanned aerial vehicles (UAVs) or autonomous underwater vehicles (AUVs) for search and survey operations \cite{fax2002information},\cite{olfati2007consensus}. 
Such systems are limited by their size and mobile (\textit{ad hoc}) nature that restrict computational and sensing resources, rendering distributed algorithms well-suited for coordination of these multi-agent systems \cite{mesbahi2010graph}. 
A number of works in the literature have proposed distributed coordination algorithms for multi-agent systems achieving consensus on the average of agents\textquotesingle \ initial state values. 
The authors in \cite{kempe2003gossip} introduced a novel gossip-based decentralized scheme called push-sum protocol to compute the average of the initial state values of nodes. 
The push-sum protocol has been shown to converge to the average exponentially fast.
Authors of \cite{dominguez2010coordination} have proposed a ratio-consensus protocol in which agents in a fixed topological network converge asymptotically to the ratio of the sum of initial value of the two states maintained by each agent. 
These algorithms have a significant advantage; convergence to the average value can be established where the protocol can be realized in a truly distributed manner without requiring any centralized dissemination of parameters making them suitable for plug-and-play \textit{ad hoc} networks. 
\cite{charalambous2014average} extends this approach to the case where network topology is dynamic under the condition that the union of the communication graphs at different time instants remain strongly connected infinitely often.
Convergence rate is an important performance indicator for consensus protocols \cite{xiao2004fast},\cite{olshevsky2009convergence}. As presented in \cite{olfati2004consensus}, convergence rate depends on the spectral properties of the interaction graph topology. Several researchers have endeavored to design interaction graphs amenable to faster convergence \cite{kim2005maximizing},\cite{xiao2004fast}.
A consensus based distributed optimization scheme utilizing the subgradient-push over a time-varying graph topology was presented in \cite{nedic2014distributed}. The subgradient iterations are shown to converge at a rate of $\bigO(\ln(t)/\sqrt{t})$. However, each node has to continue the subgradient iteration updates as there is no mechanism to distributively detect when the optimal solution is reached.

Above results do not provide any finite time stopping criterion for the consensus protocol. As multi-agent systems with real-time applications, require the consensus value to be used by each agent for a subsequent task or action, a finite-time distributed stopping criterion is imperative.
 Moreover, if the agents can distributedly detect the convergence  within a pre-specified tolerance of the consensus value, they will avoid running the algorithm longer than necessary and save scarce power and computational resources. 
The authors in \cite{sundaram2007finite} have presented a method to achieve the consensus value in a finite number of iterations. Here, each node can calculate the final consensus value using the minimal polynomial associated with the weight matrix in the state update iterations. However, to calculate the coefficients of the minimal polynomial each node has to run $N$ (total number of agents) different linear iterations each for at least $N+1$ time-steps. Also, every node should have enough storage and  computation abilities to handle matrix inversions and rank calculations which makes it unsuitable for applications like \textit{ad hoc} sensor networks.
 
To this end, authors in \cite{ratio_consensus_lab},\cite{yadav2007distributed},\cite{prakash2019distributed} established that the sequence of global maximum (or minimum) state value of the agents following a consensus algorithm is a monotonic sequence converging to the consensus when network topology is fixed. A distributed maximum (and minimum) protocol was proposed to propagate these global maximum (and minimum) state values to achieve finite-time consensus within a pre-specified tolerance margin. Having additional states corresponding to the global maximum and minimum values help each node to detect the progress toward consensus. Each node is able to simultaneously detect the convergence in finite time thus the consensus protocol is terminated by each node at the same iteration. Moreover the above methodology guarantees that each node will have access to the consensus value at the same time.

In this article, we show that the above approach is not directly applicable when the network topology is dynamic. Here, we propose an extension of this approach which can be applied to networks with dynamic topology. 
We introduce the concept of a "time-path" to incorporate the effect of current state value of an agent on other agents at the following time instants. We establish the existence of time-paths of finite length for all pairs of agents in the network. Leveraging the existence of time-paths, we propose a new Maximum-Minimum protocol to propagate the global maximum and minimum state values. 
Now, we briefly describe some of the applications of the proposed protocol.

\textit{Ad hoc cognitive radio networks.} In cognitive radio networks, secondary users can sense the spectrum to detect the presence of primary users. In a spectrum-sensing consensus algorithm \cite{li2009distributed}, secondary users mutually transmit and receive their states according to the real-time (dynamic) network topology, regardless of whether primary users are present. The topology is created when secondary users establish communication links with their own neighbors to locally exchange information among them. The algorithm iterations are repeatedly done until all the individual states converge toward the average of the initial value of states to make a local decision. 

\textit{Control of autonomous agents.} It is often necessary to coordinate a collection
of autonomous agents (e.g., cars or unmanned aerial vehicles). For example, consider a fleet of self-driving cars where each car can communicate with its neighboring cars. One may wish for the cars to meet a global objective such as maintaining a particular formation where the neighbors of a car can change in real-time. A distributed decision is usually needed in such situations. All the cars can distributively agree on a direction or an average speed (or both). Such a coordination model was investigated in \cite{fax2002information}.

\subsubsection*{ Statement of contribution:}
\begin{enumerate}
    \item This article presents an algorithm with a distributed stopping criterion for ratio consensus in the presence of directed switching topologies.  We augment the ratio consensus algorithm with two additional states: global maximum and global minimum of the values held by agents. We show that these values are monotonic in nature and converge to the consensus value even when the network is dynamic. The stopping criterion can be set such that if the criterion is met by an agent then it has access to the consensus value within any prespecified tolerance margin. Here, the maximum and minimum consensus based distributed stopping approach \cite{ratio_consensus_lab},\cite{yadav2007distributed} is extended to the case of time varying topologies.

   \item We provide an upper bound on the number of iterations required for a node in the network to influence all other nodes. In order to achieve this, we introduce a novel concept of time-path. We prove the existence of finite-length time-paths for every pair of nodes and present the Maximum-Minimum protocol to propagate the global maximum and minimum state values in the network. 
   
    \item The scheme proposed here is shown to be scalable  for implementation as it only requires each agent to have access to an upper bound on the number of nodes. 
    The performance of the proposed algorithm is illustrated by
    experimentally realizing a network with dynamic topology created using a NodeJS framework. Here each node is implemented as a \textit{socket.io} \cite{mulder2016node} server.
    This validates the distributed stopping criterion experimentally in the presence of switching topologies, thus providing evidence that our algorithm is indeed applicable for real-time applications.

\end{enumerate}

 The rest of the paper is organized as follows. In Section~\ref{sec2}, the basic definitions needed for subsequent development are presented. The setup for distributed averaging using ratio consensus is presented in Section~\ref{sec3}. In Section~\ref{sec4}, analytical results for distributed finite-time termination of ratio consensus in switching topology using maximum and minimum consensus algorithms have been discussed. Theoretical findings are validated with experiments presented in Section~\ref{sec5} followed by conclusions in Section~\ref{sec6}. 

\section{Definitions}\label{sec2}
In this section we present basic notions of graph theory and linear algebra which are essential for the subsequent developments. Detailed description of graph theory and linear algebra notions are available in \cite{Die06} and \cite{horn2012matrix} respectively.
\begin{definition}(Cardinality of a set)
Let $A$ be a set. The cardinality of a set $A$ denoted by $|A|$ is a measure of the number of elements of the set $A$. 
\end{definition}
\begin{definition}(Directed Graph)
A directed graph (denoted as digraph) $\G$ is a pair $(\V,\E)$ where $\V$ is a set of vertices or nodes and $\E$ is a set of edges, which are ordered subsets of two distinct elements of $\V$. If an edge from $j \in \V$ to $i \in \V$ exists then it is denoted as $(i,j)\in \E$.  
\end{definition}
\begin{definition}(Path) 
In a directed graph, a directed path from node $i$ to $j$
exists if there is a sequence of distinct directed edges of $\G$ of
the form $(k_{1},i),(k_{2},k_{1}),...,(j,k_{m}).$
\end{definition}
\begin{definition}(Strongly Connected Graph) A directed graph is strongly connected if it has a directed
path between each pair of distinct nodes $i$ and $j$.
\end{definition}
\begin{definition}(Column Stochastic Matrix) A real $n\times n$ matrix $A=[a_{ij}]$
is called a column stochastic matrix if $1 \geq a_{ij} \geq 0$ for $1\leq i,j\leq n$
and $\displaystyle \sum_{i=1}^{n}a_{ij}=1$ for $1\leq j\leq n.$ 
\end{definition}
\begin{definition}(Irreducible Matrix)
	A $N \times N$ matrix $A$ is said to be irreducible if for any $i, j \in \{1,...,N\}$, there exist $m \in \mathbb{N}$ such that $(\textbf{A}^m)(i,j) > 0$, that is, it is possible to reach any state from any other state in a finite number of hops.
\end{definition}
\begin{definition}(Primitive Matrix) A non negative matrix $A$ is primitive if it is irreducible and has only one eigenvalue of maximum modulus. 
\end{definition}
\section{Average Consensus in switching topology}\label{sec3}
In this section, the key result from \cite{dominguez2010coordination}, which enables reaching average consensus in the presence of dynamic topology is summarized. Consider a scenario where the network topology is dynamic but with a fixed set of nodes $\V$ ($|\V| = n$) i.e. at any given instant $k$, the network is described by a digraph $\G(k)=(\V,\E(k))$. Let $\Pd(k)=(p_{ij}(k))$ be the weighted adjacency matrix associated with the digraph $\G(k)$. Here $\G(k) \in \bar{\G} = \{\G_1, \G_2,\dots,\G_m\}, m\leq 2^{n^2-n}$ is the set of all possible digraphs for a given set of nodes $\V$. Here we assume that a node always has access to its own information, i.e. for any node $i \in \V, (i,i) \in \E(k)$ for all $k$. Now we present a few definitions related to dynamic topology.
 \begin{definition}(Union of digraphs)
Given a collection of digraphs $\{\G_1, \G_2,\dots,\G_m\}$ (for some $m \geq 1$) of the form $\G(k)=(\V,\E(k))$, $k = 1, 2, . . . , m$, the union of digraphs is defined as $\G(k)_{1,2,...,m} = (\V, \cup_{k=1}^m\E(k))$.
\end{definition}
\begin{definition}(In-Neighborhood at instant k) The set of in-neighbors of node $i \in \V$ at instant $k$ is denoted by $N^-(i,k) = {\{j: (i,j)\in \E(k)}\}.$ 
\end{definition}
\begin{definition}(Out-Neighborhood at instant k) The set of out-neighbors of node $i \in \V$ at instant $k$ is denoted by $N^+(i,k) = {\{j: (j,i)\in \E(k)}\}.$
\end{definition}
Each node $i \in \V$ maintains two states at time $k$, denoted by $x_{i}(k)$ (referred as numerator state of node $i$) and $y_{i}(k)$ (referred as denominator state of node $i$). Node $i$ updates its state at the $(k+1)^{th}$ iteration according to the following policy:
\begin{equation}
x_{i}(k+1)=p_{ii}(k)x_{i}(k)+\sum_{j\in\mathit{N^-(i,k)\setminus\{i\}}}p_{ij}(k)x_{j}(k) \label{consensus_num}
\end{equation}
\begin{equation}
y_{i}(k+1)=p_{ii}(k)y_{i}(k)+\sum_{j\in\mathit{N^-(i,k)\setminus\{i\}}}p_{ij}(k)y_{j}(k),  \label{consensus_den}
\end{equation}
where $y_i(0) = 1$ for all $i \in \V$.

We consider the network with dynamic topology to satisfy the following assumptions throughout the rest of the paper.
\begin{assump}\label{ass:unionConn}
For a sequence of digraphs $\G(k)=(\V,\E(k))$, $k = 0, 1, 2,\dots,$
there exists an infinite sequence of time instants $t_0, t_1, \dots, t_m,\dots,$ where $t_0 = 0$, $0 < t_{m+1}-t_m\leq l < \infty$, $ l\ge 0,m\ge 0$, with the property that for any $m$ the union of digraphs  $\G(t_m), \G(t_m + 1), \dots, \G(t_{m+1}-1)$ is strongly connected.
\end{assump}
\begin{assump}\label{ass:colSto}
$\Pd(k)$ for all $k$ is a column-stochastic, primitive and irreducible matrix. 
\end{assump}

\begin{theorem}\label{thm:avg_con}
Consider a sequence of digraphs of the form $\G(k)=(\V,\E(k))$, $k = 0, 1, 2,\dots$ satisfying Assumption~\ref{ass:unionConn} and Assumption~\ref{ass:colSto}. With the update rule (\ref{consensus_num}) and (\ref{consensus_den}), the ratio $\frac{x_{i}(k)}{y_i(k)}$ asymptotically converges to $\frac{{\sum_{i=1}^{n}}x_{i}(0)}{n}$ for all $i=1,...,n$, that is, the ratio of the numerator and denominator states converge to the average of the initial conditions of $x_i$ variables (referred to as ratio consensus). 
\end{theorem}
\textbf{Proof. }
See \cite{dominguez2010coordination} for proof.
\qed
\section{Distributed Finite-Time Termination}\label{sec4}
In this section, the definitions and convergence of maximum and minimum consensus algorithms are established. Subsequently, a finite-time termination criterion for average consensus in the case of switching topology is developed based on these algorithms. Let us consider the maximum and minimum value of the ratio of consensus protocols given by (\ref{consensus_num}) and (\ref{consensus_den}) over all nodes at any time instant $k$ be given by,
\begin{equation}
    M(k):=\underset{i\in {\V}}{\max} \frac{x_{i}(k)}{y_{i}(k)}, y_j(k) \neq 0, j \in \V, \label{max_ratio}
\end{equation}
\begin{equation}
     m(k):=\underset{i \in {\V}}{\min} \frac{x_{i}(k)}{y_{i}(k)}, y_j(k) \neq 0, j \in \V \label{min_ratio}
\end{equation}
    
The following Lemma shows that the ratio of states at each node stays within the maximum and minimum for subsequent iterations.
\begin{lemma}\label{lem:MXPmono}
Consider the ratio consensus protocol of (\ref{consensus_num}) and (\ref{consensus_den}). Let Assumption~\ref{ass:unionConn} and Assumption~\ref{ass:colSto} hold. Then for all time instants $k^{'}\geq k$ and for all $i\in \V,$ 
\begin{align*}
    m(k) \leq \frac{x_{i}(k^{'})}{y_{i}(k^{'})} \leq M(k).
\end{align*}
\end{lemma}
 \textbf{Proof. }
We first prove the inequality for $M(k)$ using induction. By definition of $M(k),$ for $k^{'}=k,$ the proof is trivial.  
 Suppose it is asserted that for $k^{'}=k+l, l \geq 1$, $\frac{x_{i}(k+l)}{y_{i}(k+l)}\leq M(k)$ for all $i\in \V.$ Then we have,
 \begin{equation*}
 \frac{x_{i}(k+l+1)}{y_{i}(k+l+1)}=\frac{\sum\limits_{j\in N^-(i,k+l)} p_{ij}(k+l)x_{j}(k+l)}{\sum\limits_{j\in N^-(i,k+l)} p_{ij}(k+l)y_{j}(k+l)}
 \end{equation*}
 \begin{equation*}
 = \frac{p_{ii}(k+l)\frac{x_{i}(k+l)}{y_{i}(k+l)} +\sum\limits_{j\in N^-(i,k+l)\setminus\{i\}} p_{ij}(k+l)\frac{x_{j}(k+l)}{y_{i}(k+l)}}{p_{ii}(k+l)+\sum\limits_{j\in N^-(i,k+l)\setminus\{i\}} p_{ij}(k+l)\frac{y_{j}(k+l)}{y_{i}(k+l)}}.
 \end{equation*}
It follows from the inductive assumption that,
 \begin{equation*}
 \frac{x_{i}(k+l+1)}{y_{i}(k+l+1)}\leq
 \frac{p_{ii}(k+l)M(k)+\sum\limits_{j\in N^-(i,k+l)\setminus\{i\}} p_{ij}(k+l)M(k)\frac{y_{j}(k+l)}{y_{i}(k+l)}}{p_{ii}(k+l)+\sum\limits_{j\in N^-(i,k+l)\setminus\{i\}} p_{ij}(k+l)\frac{y_{j}(k+l)}{y_{i}(k+l)}}
 = M(k).
 \end{equation*}

 Therefore, $\frac{x_{i}(k+l+1)}{y_{i}(k+l+1)}\leq M(k)$ for all $i\in \V.$ Other inequality is similar and is left to the reader.
 \qed

 Next Lemma strengthens the result of Lemma~\ref{lem:MXPmono} to a strict inequality.
 \begin{lemma}\label{lem:MXPsMono}
Consider the ratio consensus protocol of (\ref{consensus_num}) and (\ref{consensus_den}) where the initial time instant is $k$. Let Assumption~\ref{ass:unionConn} and Assumption~\ref{ass:colSto} hold.  Let $M(k)$ and $m(k)$ be as in (\ref{max_ratio}) and (\ref{min_ratio}). Let $i$ be a node such that $\frac{x_{i}(k^{'})}{y_{i}(k^{'})}<M(k)$ and let $j$ be a node such that $\frac{x_{j}(k^{'})}{y_{j}(k^{'})}>m(k)$ for some time instant $k^{'}\geq k$. Then for all time instants $k^{''}\geq k^{'}, \frac{x_{i}(k^{''})}{y_{i}(k^{''})}<M(k)$ and $\frac{x_{j}(k^{''})}{y_{j}(k^{''})}>m(k).$ 
 \end{lemma}
 \textbf{Proof. }
The proof is based on induction and follows similarly to the proof of Lemma~\ref{lem:MXPmono} and is left to the reader.
 \qed
 
The following definition and Lemmas introduce the concept of a time-path and derive a bound on number of iterations required for one node to access information of any other node in the network.
\begin{definition}(Time-path)
In the case of switching topology, a time-path of length $l$ at time $t$ from node $i \in V$ to
$j \in V$ is a sequence of nodes $k_1,k_2,\dots,k_{l-1}$ such that $(k_{1},i) \in \E(t),(k_{2},k_{1}) \in \E(t+1),\dots ,(j,k_{l-1})\in \mathcal{E}(t+l-1)$. In other words, node $j$ has access to node $i$'s information after $l$ time steps through the nodes $k_1,k_2,\dots,k_{l-1}$ starting at time $t$.
\end{definition}
 \begin{lemma}\label{lem:rangeOFi}
Consider a network where Assumption~\ref{ass:unionConn} and Assumption~\ref{ass:colSto} hold along with added constraint that digraphs $\G(k)$ for all $k \in \mathbb{N}$ are strongly connected. Then for any node $i \in \V$,  let 
 \begin{align*}
     R_i(k,t):=\bigcup\limits_{m \in R_i(k,t-1)} \{l : l \in N^+(m,k+t-1)\},
 \end{align*}
 with, $R_i(k,0) = \{ i \}, $ $|R_i(k,0)|=1 $ and $N^+(m,k)$ is the out-neighborhood of node $m$ at instant $k$. The following hold:
\begin{enumerate}
\item $R_i(k,t-1)\subset R_i(k,t)$ for all $t=1,2\ldots.$
\item $R_i(k,n-1)=\mathcal{V}.$
\end{enumerate}
 
 \end{lemma}
 \textbf{Proof. }
 Here, $R_i(k,t+1)$ is the union of out-neighborhoods of all the elements of $R_i(k,t)$ at instant $t$.  Clearly, $R_i(k,t-1) \subseteq$ $R_i(k,t)$. Indeed, $l \in R_i(k,t-1)$ implies $l \in N^+(l,k+t-1)$ which in turn implies, $l \in R_i(k,t)$. 
 
 \noindent Now for $t=1$, 
 \[R_i(k,1) = \{ m_1 : m_1 \in N^+(i,k) \} = N^+(i,k).\]
 If $R_i(k,1) = \V$, the claim is proven. 
 Suppose $R_i(k,1) \subset \V$. Note that at time instant $k+1$, the graph is strongly connected. Thus, there exists an outgoing edge between the set of nodes $R_i(k,1)$ and $\V \setminus R_i(k,1)$. That is there exists $m_2 \in \V \setminus R_i(k,1)$ such that $m_2 \in N^+(m_1, k+1)$ for some $m_1 \in R_i(k,1)$. Thus, $m_2 \in R_i(k,2).$ It follows that $m_2 \in R_i(k,2) \setminus R_i(k,1)$ which implies that 
\[|R_i(k,2) \setminus R_i(k,1)| \geq 1.\]
Now, as $R_i(k,1) \subset R_i(k,2)$ we have, 
\[|R_i(k,2)| = |R_i(k,1)| + |R_i(k,2) \setminus R_i(k,1)|\]
 which implies $|R_i(k,2)| \geq |R_i(k,1)| + 1.$
Note here $R_i(k,1)$ is a proper subset of $R_i(k,2).$ Following in the same manner we get, 
\[ |R_i(k,t)| \geq |R_i(k,t-1)| + 1 \geq |R_i(k,t-2)| +2 \geq \dots \geq |R_i(k,0)| + t,\]
and thus $|R_i(k,t)| \geq t + 1.$ For $t=n-1$, $|R_i(k,n-1)| \geq n $. However $R_i(k,t) \subseteq \V$ and thus $|R_i(k,t)| \leq n$ for all $t$. Thus, $|R_i(k,n-1)| = n $ implying $R_i(k,n-1) = \V$. Therefore, the claim is true. 
 \qed
 \begin{lemma}\label{lem:timePath}
 Consider the assumptions of Lemma~\ref{lem:rangeOFi}. Then for any two nodes $i,j \in \V$ and at any time instant $k$, there exist a time-path of length $s$ from $i$ to $j$ with $s\leq n-1$. In other words, $j$ has access to information of $i$ in $s$ number of time steps. 
 \end{lemma}
 \textbf{Proof. }
It is to be noted here that $R_i(k,t)$ is the set of nodes influenced by node $i$'s current state within next $t$ time steps.
Now, we show the existence of a time-path. We have $R_i(k,0) \subset R_i(k,1) \subset R_i(k,2) \subset \dots \subset R_i(k,s) = \V $. Let $s$ be the smallest time step such that  $j \in R_i(k,s)$. From Lemma~\ref{lem:rangeOFi}, $s \leq n-1$. As, $j\in R_i(k,s),$ there exists $m_{s-1} \in R_i(k,s-1)$ such that $j \in N^+(m_{s-1}, k+s-1).$ 
As $m_{s-1}\in R_i(k,s-1)$ it follows that there exists $m_{s-2} \in R_i(k,s-2)$ such that $m_{s-1} \in N^+(m_{s-2}, k+s-2).$ Choosing $m_i$'s in this manner we get $m_{s-2} \in R_i(k,s-2), m_{s-3} \in R_i(k,s-3)$, \dots $m_1 \in R_i(k,1)$ such that $m_1 \in N^+(i,k)$. Therefore, there exists a time-path $(m_1,i),(m_2,m_1),\dots ,(j,m_{s-1})$ with $s \leq n-1$. 
 \qed
 
Using Lemma~\ref{lem:rangeOFi} and Lemma~\ref{lem:timePath}, we next provide a sampling of the $M(k)$  and $m(k)$ such that the resulting sub-sequences are strictly monotonic and converge to the average of the initial conditions of $x_i$ variables.
 
 \begin{lemma}\label{lem:strictMonoAll}
 Consider the ratio consensus protocol of (\ref{consensus_num}) and (\ref{consensus_den}) with the assumptions of Lemma~\ref{lem:rangeOFi}. Let $M(k)$ and $m(k)$ be as in (\ref{max_ratio}) and (\ref{min_ratio}) such that $m(k)<M(k)$ where initial time instant is $k$. Then for all $k^{'}\geq k+n'$ and for all $i \in \V,$
 \begin{align}
     m(k) < \frac{x_{i}(k^{'})}{y_{i}(k^{'})} < M(k),
\end{align}
where $n'$ is an upper bound on $n-1$.
 \end{lemma}
 \textbf{Proof. }
As $m(k)< M(k)$ it follows that there exists a node $i \in \V$ such that $\frac{x_{i}(k)}{y_{i}(k)}<M(k).$ Let $j\in \V$ be an arbitrary node, then from Lemma~\ref{lem:timePath} there exist a time-path of length $l$ from node $i$ to node $j$ at instant $k$ where $l \leq n-1 \leq n'$. Let this path be denoted as $(m_{1},i), (m_{2},m_{1}),..., (j,m_{l-1}).$ Then,
 \begin{equation*}
 \frac{x_{m_{1}}(k+1)}{y_{m_{1}}(k+1)}
 =\frac{p_{m_{1}i}(k)x_{i}(k)+\sum\limits_{u\in N^-(m_{1},k) \setminus \{i\}} p_{m_{1}u}(k)x_{u}(k)}{p_{m_{1}i}(k)y_{i}(k)+\sum\limits_{u\in N^-(m_{1},k)\setminus \{i\}} p_{m_{1}u}(k)y_{u}(k)}
 \end{equation*}
 \begin{equation*}
 =\frac{p_{m_{1}i}(k)\frac{x_{i}(k)}{y_{i}(k)}+\sum\limits_{u\in N^-(m_{1},k)\setminus \{i\}} p_{m_{1}u}(k)\frac{x_{u}(k)}{y_{i}(k)}}{p_{m_{1}i}(k)+\sum\limits_{u\in N^-(m_{1},k)\setminus \{i\}} p_{m_{1}u}(k)\frac{y_{u}(k)}{y_{i}(k)}}
 \end{equation*}
 \begin{equation*}
 <\frac{p_{m_{1}i}(k)M(k)+\sum\limits_{u\in N^-(m_{1},k)\setminus \{i\}} p_{m_{1}u}(k)\frac{x_{u}(k)}{y_{i}(k)}}{p_{m_{1}i}(k)+\sum\limits_{u\in N^-(m_{1},k)\setminus \{i\}} p_{m_{1}u}(k)\frac{y_{u}(k)}{y_{i}(k)}}.
 \end{equation*}
 By definition of $M(k), \frac{x_{u}(k)}{y_{u}(k)}\leq M(k)$ for all $u \in \V$, thus $x_{u}(k)\leq y_{u}(k)M(k).$ It follows that,
 \begin{equation*}
 \frac{x_{m_{1}}(k+1)}{y_{m_{1}}(k+1)}<
 \frac{p_{m_{1}i}(k)M(k)+\sum\limits_{u\in N^-(m_{1},k)\setminus \{i\}} p_{m_{1}u}(k)M(k)\frac{y_{u}(k)}{y_{i}(k)}}{p_{m_{1}i}(k)+\sum\limits_{u\in N^-(m_{1},k)\setminus \{i\}} p_{m_{1}u}(k)\frac{y_{u}(k)}{y_{i}(k)}}
 =M(k). 
 \end{equation*}
Thus, $\frac{x_{m_{1}}(k+1)}{y_{m_{1}}(k+1)}<M(k).$  Therefore, from Lemma~\ref{lem:MXPsMono} it follows that for all $k^{'}\geq k+1$, $\frac{x_{m_{1}}(k^{'})}{y_{m_{1}}(k^{'})}< M(k).$ 
 Similarly, if $k^{'}\geq k+2$, then $\frac{x_{m_{2}}(k^{'})}{y_{m_{2}}(k^{'})}<M(k)$ and that for all $k^{'}\geq k+l, \frac{x_{j}(k^{'})}{y_{j}(k^{'})}<M(k).$ Note that since $n'\geq n-1 \geq l,$ it follows that $k+n'\geq k+l.$ The condition $k'\geq k+n'$ is independent of the index $j$ and where the node $j$ was chosen arbitrarily. Thus, it can be concluded that $\frac{x_i(k')}{y_i(k')}<M(k)$ for all $k'>k+n'$ and for all $i\in \V.$ This completes the proof for $M(k).$ The other inequality for $m(k)$ can be proven similarly and is left to the reader.
 \qed
 \begin{remark}
Note that from Lemma~\ref{lem:strictMonoAll}, after a finite number of iterations given by $n'$, all ratios of the nodal states under (\ref{consensus_num}) and (\ref{consensus_den}) become strictly less than the maximum value of the ratio in network in the past and strictly greater than the
minimum value of the ratio in the network in the past. 
 \end{remark}
 The following theorem shows that after a finite time, the maximum value of the ratio in the network decreases and the minimum value of the ratio in the network
increases. 
 \begin{theorem}\label{thm:mxpMNPSeq}
 Consider the ratio consensus protocol of (\ref{consensus_num}) and (\ref{consensus_den}) with the assumptions of Lemma~\ref{lem:rangeOFi} and the initial ratio vector being 
 \[r(un'):=\Big[\frac{x_1(un')}{y_1(un')}\ \frac{x_2(un')}{y_2(un')} \ldots \frac{x_N(un')}{y_N(un')}\Big]\]
 such that $\min r(un')< \max r(un')$, where, $u=0, 1, 2,...$. Then, $M((u+1)n')<M(un')$ and $m((u+1)n')>m(un').$  
 \end{theorem}
 \textbf{Proof. }
 Let $k=un'$, then using Lemma~\ref{lem:strictMonoAll}, it follows that for $k^{'}\geq (u+1)n', \frac{x_{i}(k^{'})}{y_{i}(k^{'})}<M(un')$ for all $i \in \V.$ Hence, $M((u+1)n'):=\underset{i \in {\V}}{\max} \frac{x_{i}((u+1)n')}{y_{i}((u+1)n')}<M(un').$ This completes the proof of the inequality involving $M(un')$. The other inequality involving $m(un')$ can be proved similarly and is left to the reader.
 \qed
 
 \begin{theorem}\label{thm:mxpMNPCvg}
 Consider the ratio consensus protocol of (\ref{consensus_num}) and (\ref{consensus_den}) with the assumptions of Lemma~\ref{lem:rangeOFi}. Then, 
 \begin{align*}
     \lim_{u\to\infty}  M(un')&=\frac{\sum_{j=1}^{n} x_{j}(0)}{n}\ \text{ and}\\
     \lim_{u\to\infty}  m(un')&=\frac{\sum_{j=1}^{n} x_{j}(0)}{n}.
 \end{align*}
 \end{theorem}
 \textbf{Proof. }
 This result follows directly from Theorem~\ref{thm:avg_con} and Theorem~\ref{thm:mxpMNPSeq} and is left to the reader.
 \qed
 \begin{corollary}\label{cor:mxpMNP}
 Consider the ratio consensus protocol of (\ref{consensus_num}) and (\ref{consensus_den}) with the assumptions of Lemma~\ref{lem:rangeOFi}. Then, 
 \[\lim_{u\to\infty}  M(un')-m(un')=0.\]
 \end{corollary}
 \textbf{Proof. }
 The proof follows from Theorem~\ref{thm:mxpMNPCvg} and is left to the reader.
 \qed
 
We next present Maximum-Minimum consensus protocol in the case of dynamic topology and use the preceding theorems to design a finite time termination criterion.
\begin{subsection}{Maximum-Minimum Consensus Protocol}
The Maximum and Minimum Consensus Protocol denoted by MXP and MNP computes
the maximum and minimum of the given initial node conditions
\begin{align*}
   z(0) = [z_{1}(0) \ z_{2}(0)\ ... \ z_{n}(0)]^T,\\
   w(0) = [w_{1}(0)\ w_{2}(0)\ ... \ w_{n}(0)]^T
\end{align*}
in a distributed manner respectively. It takes $z(0)$ and $w(0)$ as an input and generates a sequence of node values based on the following update rules for node $i$,
\begin{align}
    z_{i}(k+1)&= \underset{j\in N^-(i,k)}{\max} z_{j}(k), \label{eq:MXP}\\
    w_{i}(k+1)&= \underset{j\in N^-(i,k)}{\min} w_{j}(k). \label{eq:MNP}
\end{align}
\begin{proposition}\label{prop:MXP}
MXP protocol given by (\ref{eq:MXP}) converges to $ \underset{j \in \V}{\max}$ $z_{j}(0)$ in finite time $k \leq n'$ for any $n'\geq n-1.$
\end{proposition}
\textbf{Proof. }
Let $m$ be a node with state value at $z_m(0) = \underset{j \in \V}{\max}$ $z_{j}(0)$. Then at $k=1$, all nodes connected to $m$ at instant $k=0$ will have the maximum value $z_m(0)$. Then, from the definition of $R_m(0,k)$ in Lemma~\ref{lem:timePath}, if $m_1 \in R_m(0,1)$ then $z_{m_1}(1)=z_m(0).$ Similarly, at $k=2$, all the nodes connected to the elements of the set $R_m(0,1)$ at instant $k=1$ will have the maximum value, that is, if $m_2 \in R_m(0,2)$ then $z_{m_2}(2)=z_m(0)$ and so on. At $k=n-1$,  if $m_{n-1} \in R_m(0,n-1)$ then $z_{m_{n-1}}(n-1)=z_m(0)$. As $R_m(0,n-1)=\V$, we have for all $j \in \V$, $z_j(n-1)=z_m(0).$ Thus, at instant $k=n-1$ all nodes will have the maximum value. As $n'\geq n-1$, $z_{j}(k)$ converges to $ \underset{j \in \V}{\max}$ $z_{j}(0)$ in finite time $k \leq n'.$
\qed
\begin{proposition}\label{prop:MNP}
MNP protocol given by (\ref{eq:MNP}) converges to $ \underset{j \in \V}{\min}$ $w_{j}(0)$ in finite time $k \leq n'$ for any $n'\geq n-1.$
\end{proposition}
\textbf{Proof. }
Similar to the proof of Proposition~\ref{prop:MXP}
\qed
\end{subsection} 
\begin{subsection}{Distributed Finite-Time Termination Algorithm for Ratio Consensus}
Here, we propose an algorithm using the above MXP-MNP protocol which allows each node to simultaneously detect the convergence of the ratio consensus within a pre-specified threshold $\rho$. 
In the proposed algorithm, the initial conditions for the MXP-MNP protocol are set as ratio of the initial values held by the nodes. 
\begin{definition}(Epoch)
$u^{th}$ epoch is defined as the  state update iteration at the instant $un'$ for any positive integer $u$.
\end{definition}

The MXP-MNP protocol is re-initialized at every $u^{th}$ epoch that is $k = un'$, where $u = 1,2,...$, with $z(un') = \frac{x(un')}{y(un')}$ and $w(un') = \frac{x(un')}{y(un')}$ respectively. We define $\bar{\alpha}_i(un') := \max z(un')=M(un')$, $\underline{\alpha}_i(un') = \min w(un')=m(un')$ and $\beta_i(un') := \bar{\alpha}_i - \underline{\alpha}_i= M(un')-m(un')$. 
Each node compares $\beta_i$ with $\rho$ at every epoch and if $\beta_i <\rho$ then it terminates the consensus protocol. Details of this scheme are given in Algorithm~\ref{alg:algo1}.

\begin{algorithm}[h]
    \SetKwBlock{Input}{Input:}{}
    \SetKwBlock{Initialize}{Initialize:}{}
    \SetKwBlock{Repeat}{Repeat:}{}
    \Input{$x_j(0),y_j(0)=1, j\in N^-(i,0), \rho$ \tcp*{Initial condition}}
    \Initialize{$k := 0$; $z_i := x_i(0)/y_i(0)$; $w_i := x_i(0)/y_i(0)$; $u := 1$;
      }
    \Repeat {
    \tcc{ratio consensus updates of node $i$ given by (\ref{consensus_num}) and (\ref{consensus_den})}
        $x_{i}(k+1) := \sum\limits_{j\in \mathit{N^-(i,k)}}p_{ij}(k)x_{j}(k)$; \
        $y_{i}(k+1) := \sum\limits_{j\in \mathit{N^-(i,k)}}p_{ij}(k)y_{j}(k)$;\\
        { 
            \tcc{global max-min updates of node $i$ given by (\ref{eq:MXP}) and (\ref{eq:MNP}) respectively}
            $\displaystyle z_i := \max_{j \in N^-(i,k)}z_j$; \
            $w_i := \displaystyle \min_{j \in N^-(i,k)}w_i$;\\
        }
        \If {$ k=  un'$} {
            $\bar{\alpha}_i := z_i$; $\underline{\alpha}_i := w_i$; $\beta_i := \bar{\alpha}_i - \underline{\alpha}_i$;\\
        \uIf {$\beta_i < \rho$ } {\textbf{break} \tcp*{stop $x_i, y_i, z_i$ and $w_i$ updates}
        }
        \uElse {set $z_{i} = x_{i}(un')/y_i(un')$; \ $w_{i} =x_{i}(un')/y_{i}(un')$;\\
        $u = u+1$;\\
        }
        } 
        $k=k+1$;\\
        }\vspace{0.1in}
        \caption{Finite-time termination of ratio consensus for switching topology (at each node $i \in \V$)}
        \label{alg:algo1}
\end{algorithm}

\begin{theorem}
Algorithm~\ref{alg:algo1} converges in finite-time simultaneously at each node.
\end{theorem}
\textbf{Proof. }
From Corollary~\ref{cor:mxpMNP}, it follows that $M(un')-m(un')\rightarrow0$ as $u\rightarrow\infty.$ Thus, for any given $\rho>0,$ there exists an integer $t(\rho)$ such that for all $u\geq t(\rho),$ $|M(un')-m(un')|<\rho$ for all nodes in the network. As each node has access to $M(un')$ and $m(un')$, convergence is detected simultaneously by each node at the same iteration.
\qed

\begin{remark}\label{rem}
Notice that using the above protocol, the global maximum and minimum values at any instant $k$ are available to each node at instant $k+n'$.
Further, the only global parameter needed for Algorithm~\ref{alg:algo1} is the knowledge of number of nodes of the network. However, it should be noted that each node does not need to know the actual number of nodes but some upper bound. In most applications, an upper bound on the number of nodes is readily available.
\end{remark}

\begin{figure}[h]
    \centering
    \begin{tabular}{cc}
       \includegraphics[scale=0.3, trim={0cm 0cm 0cm 0cm}, clip] {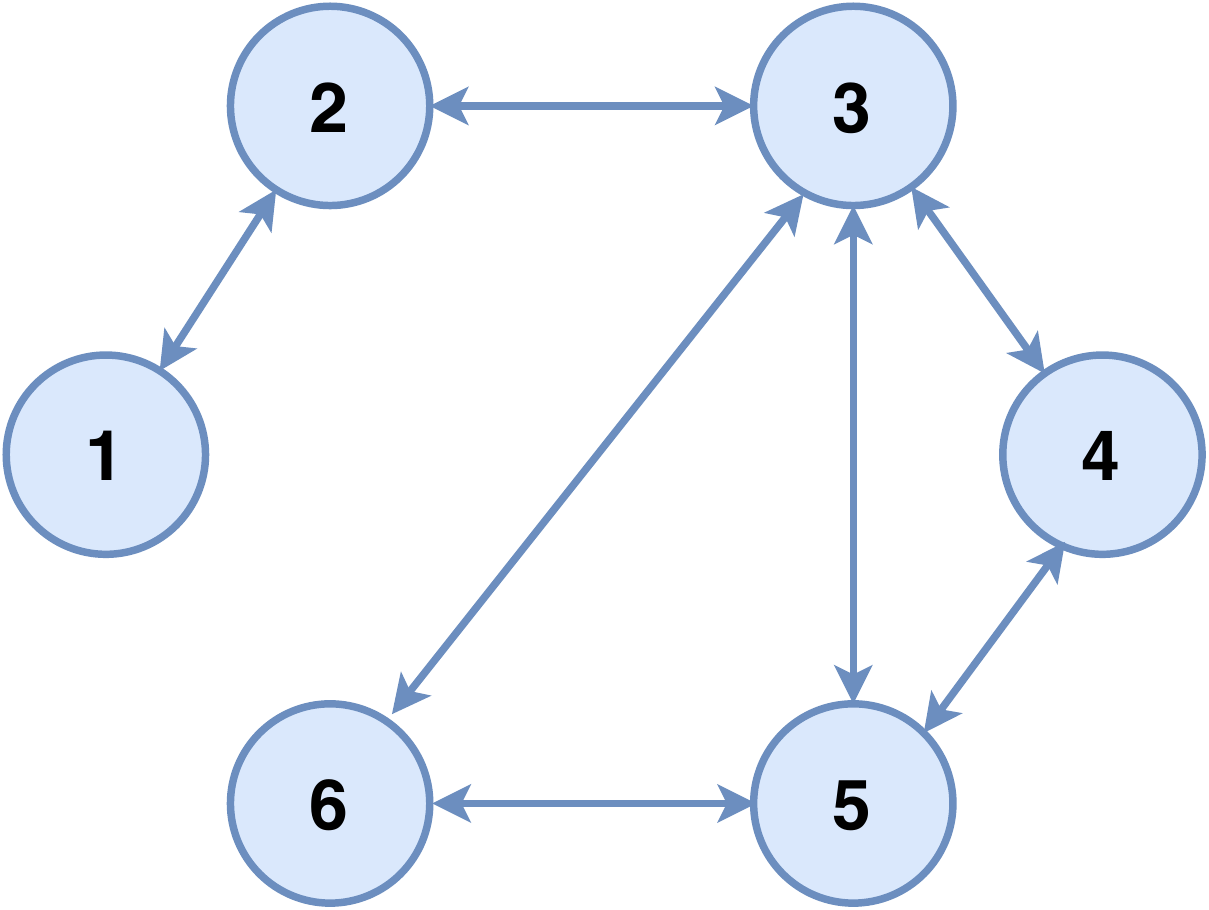}  &  \includegraphics[scale=0.3,trim={0cm 0cm 0cm 0cm},clip]{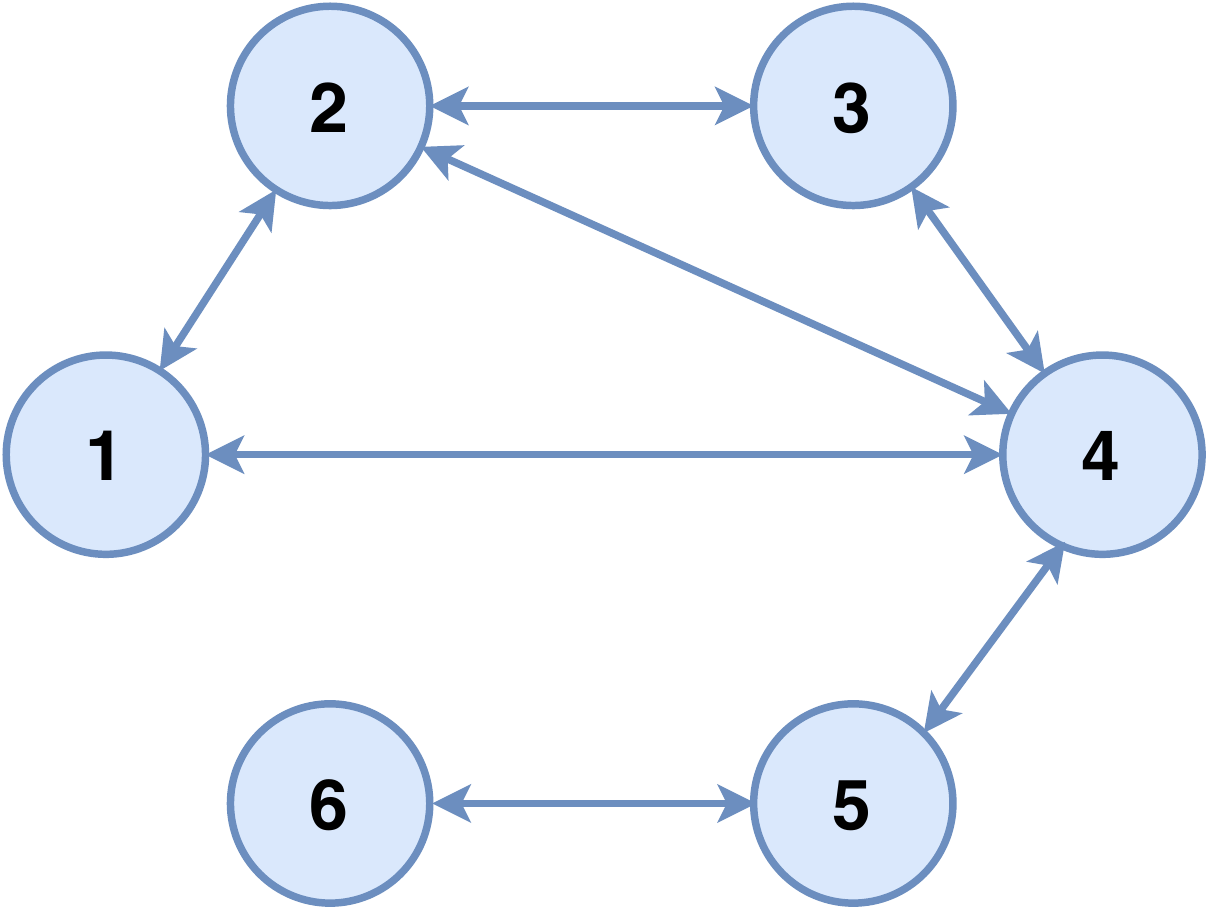}\\
       ($\G_1$) & ($\G_2$)
    \end{tabular}
      \caption{Both $\G_1$ and $\G_2$ have diameter $3$.}
    \label{fig:counterExGraphs}
\end{figure}

\begin{table}[b]
    \centering
    \begin{tabular}{|c|c|c|c|}
        \hline
         \textit{Iteration} & \textit{Topology} &  \textit{Range of node $1$}\\
         \hline
         $t=1$ & $\G_1$ & $R_1(k,1)=\{1,2\}$\\
         $t=2$ & $\G_1$ & $R_1(k,1)=\{1,2,3\}$\\
         $t=3$ & $\G_2$ & $R_1(k,1)=\{1,2,3,4\}$\\
         $t=4$ & $\G_2$ & $R_1(k,1)=\{1,2,3,4,5\}$\\
         $t=5$ & $\G_2$ & $R_1(k,1)=\{1,2,3,4,5,6\}$\\
         \hline
    \end{tabular}\vspace{0.1in}
    \caption{Network topology and range ($R_1(k,t)$) of node $1$ starting at instant $k$ for iterations $t=1,2,3,4,5$.}
    \label{tab:switchingPattern}
\end{table}
This result is a non trivial extension of \cite{ratio_consensus_lab},\cite{yadav2007distributed} as these considered only static network where the finite time consensus is based on knowledge of upper bound on graph diameter. There it was derived that within $D_{max}$ iterations every node has access to global maximum and minimum where $D_{max}$ is an upper bound on the graph diameter.
This is not applicable to a network with dynamic topology. The following counter example highlights the case where an upper bound on maximum diameter ($D_{max}=\{max(D(\G));\G \in \bar{\G}\}$ where $D(\G)$ is diameter of graph $\G$) does not provide access to global maximum and minimum to all the nodes of the network. Let us consider a network of 6 nodes where the network topology can be either of undirected graphs $\G_1$ or $\G_2$ (see Figure~\ref{fig:counterExGraphs}), that is $\bar{\G}=\{\G_1,\G_2\}$, with switching as described in Table~\ref{tab:switchingPattern}. Here the diameter of both graphs $D=3$. Range (as defined in Lemma~\ref{lem:rangeOFi}) of node 1 starting from any instant $k$ is also shown in  Table~\ref{tab:switchingPattern}. 
It can be clearly observed that it takes $5$ iterations for range to contain all the nodes which is more than the bound on maximum diameter ($D_{max}=3$). In other words, it requires at least $5\ (=n-1)$ iterations for any node to receive information from node $1$ which follows directly from Lemma~\ref{lem:timePath}.
Therefore, if $n'$ is set as $D_{max}$ in Proposition~\ref{prop:MXP} and Proposition~\ref{prop:MNP}, $z_i$ and $w_i$ for a node $i$ will not converge to global maximum and minimum respectively and rather converge to local maximum and minimum of the neighborhood defined by $R_i(k,D_{max})$. This will lead to two problems:
\begin{itemize}
    \item Theorem~\ref{thm:mxpMNPSeq} gives strict monotonicity of global maximum/minimum and does not extend to local maximum/minimum. Thus, the resulting sequences may not be monotonic.
    \item Some nodes in Algorithm~\ref{alg:algo1} can detect local convergence (when the maximum and minimum of the neighborhood defined by $R_i(k,D_{max})$ are within tolerance) and terminate well before actual global convergence.
\end{itemize}
Above analysis is further bolstered by experimental results in Section~\ref{sec5}.

\end{subsection}
\section{Experimental Results}\label{sec5}
\begin{figure}[t]
    \centering
    \begin{tabular}{ccc}
      \includegraphics[scale=0.3, trim={5cm 9cm 4.5cm 9cm}, clip] {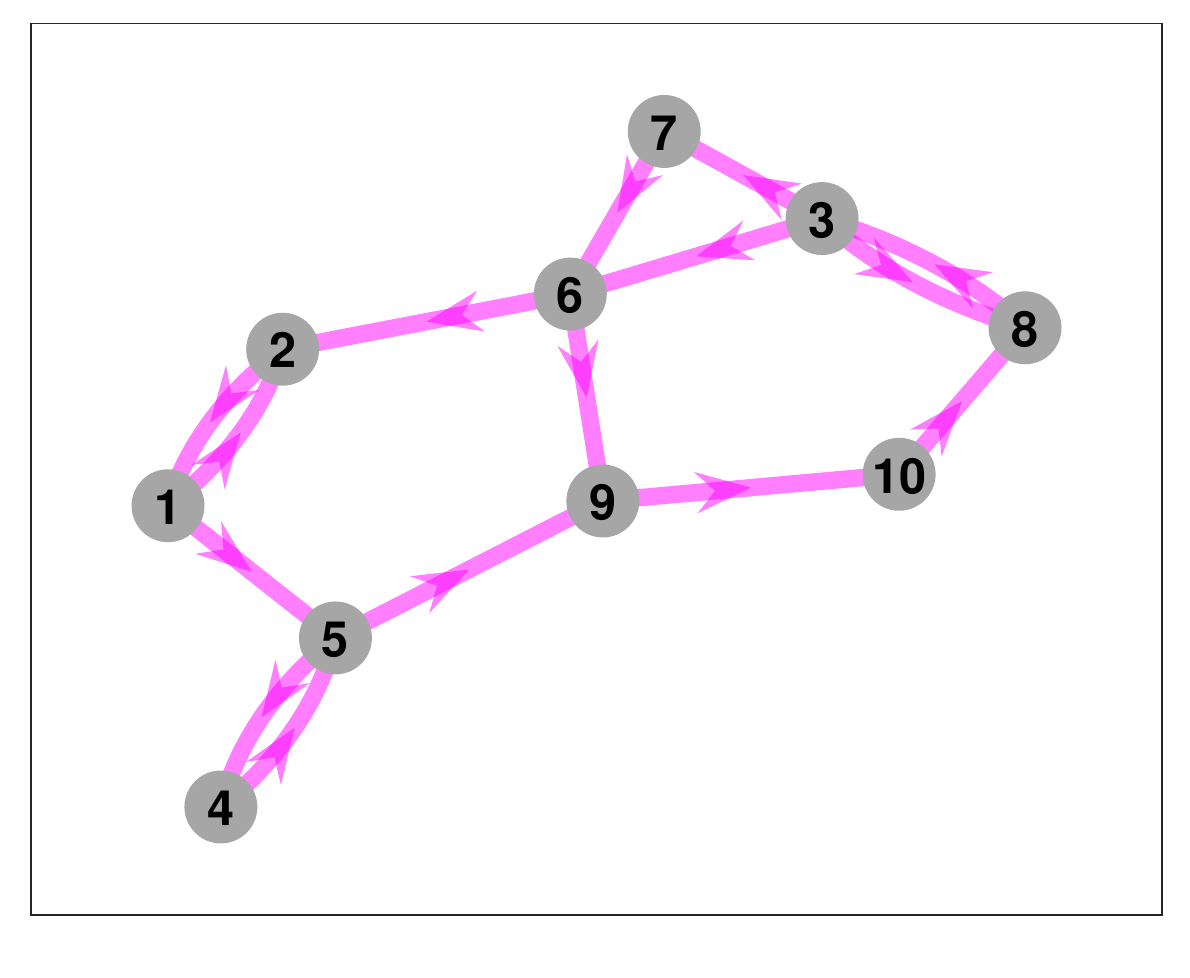}  & \hspace{-0.3cm}  \includegraphics[scale=0.3,trim={5cm 9cm 4.5cm 9cm}, clip] {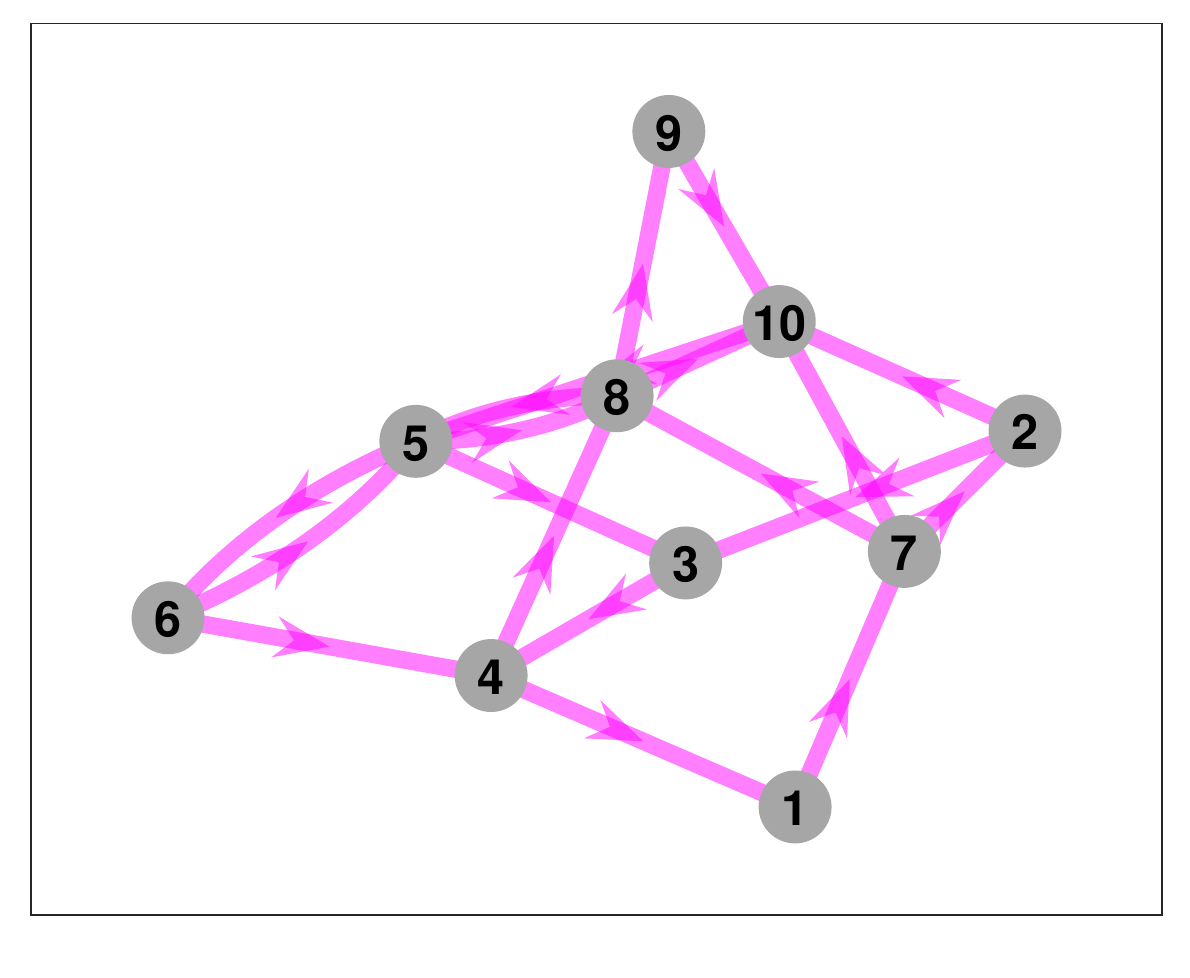} & 
      \hspace{-0.3cm}  \includegraphics[scale=0.3,trim={5cm 9cm 4.5cm 9cm}, clip] {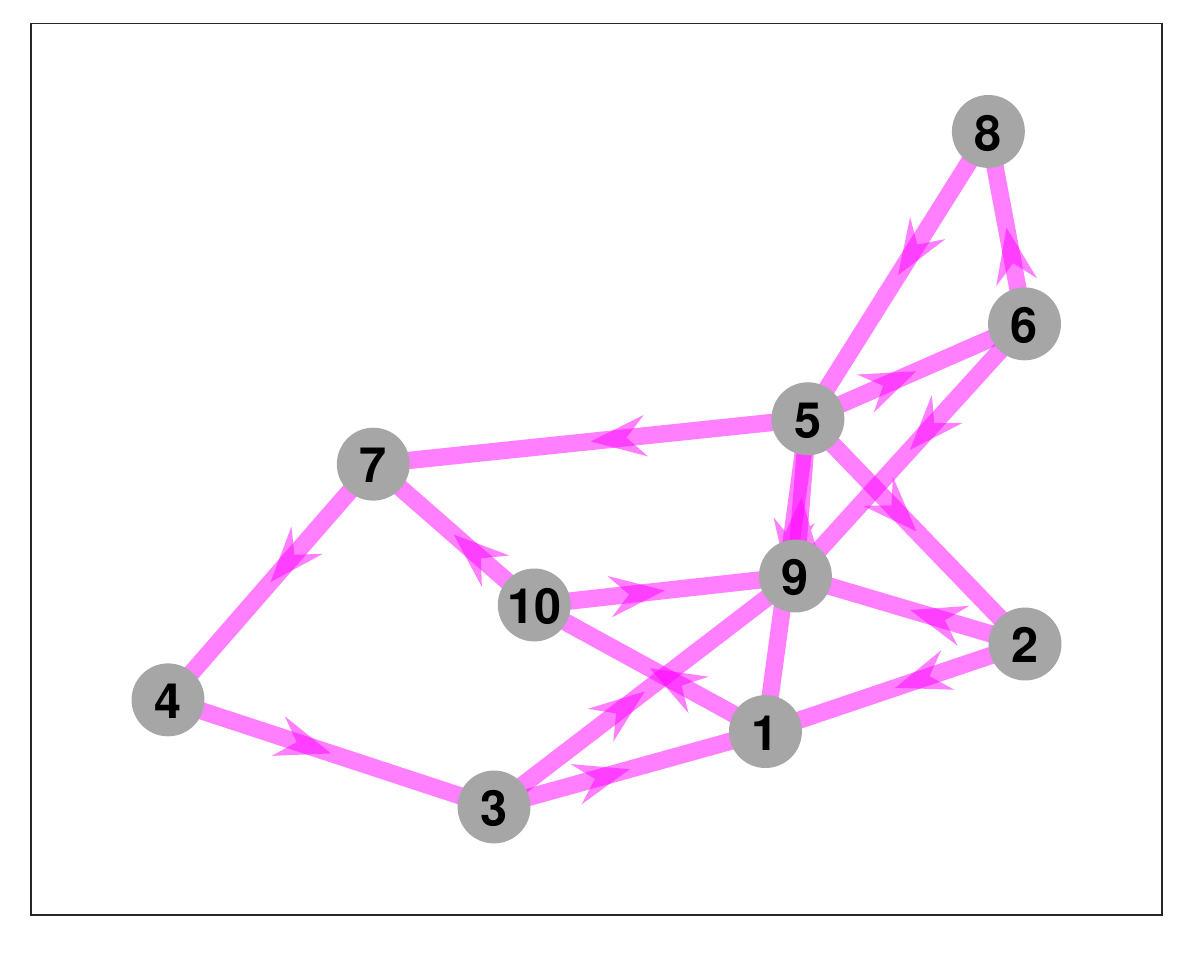}\\
      (a) & (b) & (c)
    \end{tabular}
      \caption{Topology at time instants (a) $k=9$, (b) $k=18$ and (c) $k=27$.}
    \label{fig:switchingTop}
\end{figure}
\begin{figure}[h]
    \centering
    \includegraphics[scale=0.6,trim={3cm 8.5cm 3cm 8.5cm},clip]{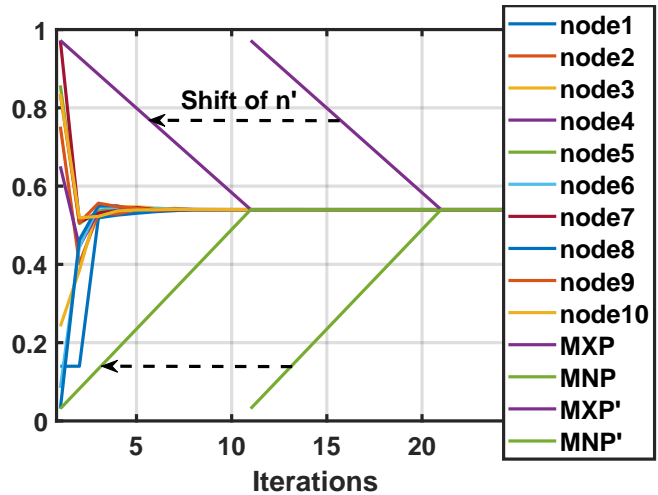}
    \caption{Finite-time termination of ratio consensus on a $10$ node dynamic network in $21$ iterations with $\rho = 0.01$.}
    \label{fig:finiteDetection}
\end{figure}

Consider a network of $10$ agents with switching topology represented by a digraph chosen at random from a set of 100 strongly connected digraphs for every time instant. Three instances of these digraphs are shown in Figure~\ref{fig:switchingTop}. Here, each agent running the consensus Algorithm $1$ is implemented using a NodeJS \textit{socket.io} server. Initial condition of the numerator states are chosen at random from $(0,1)$. With $\rho = 0.01$ and $n' = 10$, Algorithm $1$ results in distributed finite-time termination of computations performed by the agents in $21$ iterations as shown in Figure~\ref{fig:finiteDetection}. 
Observe that the ratio of the nodal states are close to the average of the numerator initial conditions (consensus value). The consensus value for this instance is $0.54$. This experiment was repeated multiple times by randomly choosing digraphs for each iteration as well as the initial values of the agents. The proposed MXP-MNP protocol was able to distributively terminate the algorithm each time. As noted in Remark~\ref{rem}, the global maximum and minimum state values at $k$ are available at $n'+k$. This is shown in Figure~\ref{fig:finiteDetection} where shifting the MXP and MNP plots left by $n'$ time instants we get the MXP$^{'}$ and MNP$^{'}$ plots which coincide with global maximum and minimum.

\begin{remark}
It is worth noting that while running MXP-MNP protocol, global maximum and minimum are available to each agent after every $n'$ iterations. It entails that if all the node states were within tolerance margin at any epoch $u$, algorithm will terminate at the next epoch resulting in a delay of $n'$ iterations in detection of convergence. This delay in detection scales linearly with the number of nodes making the algorithm suitable for large networks. 
\end{remark}

\begin{figure}[h]
    \centering
    \includegraphics[scale=0.6,trim={3cm 8.5cm 3cm 8.5cm},clip]{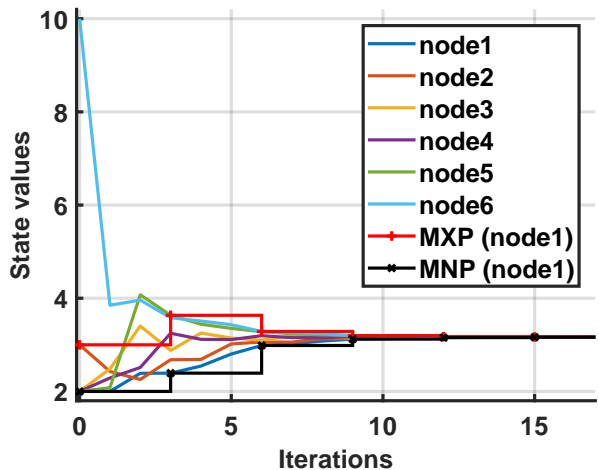}
    \caption{Non monotonicity of MXP when maximum of diameter is used as sampling.}
    \label{fig:nonMonotonicity}
\end{figure}

We next present the specific example discussed in Section~\ref{sec4} where it was shown analytically that some of the nodes will not have access to global maximum and minimum within $D_{max}$ iterations. We consider a $6$ node network where the topology is given by the sequence $\{\G_2,\G_2,\G_1,\G_2,\G_2,\G_1,\dots\}$ with $\G_1$ and $\G_2$ as defined in Figure~\ref{fig:counterExGraphs}. Initial state value of the nodes is chosen as $x(0)=[2,3, 2, 2, 2, 10]$ leading to a consensus value equal to $3.5$. As earlier, each node is implemented using a NodeJS \textit{socket.io} server. We use $n'=D_{max}=3$ to reset $z_i$ and $w_i$ in Algorithm~\ref{alg:algo1} and plot the results in Figure~\ref{fig:nonMonotonicity}. It can be clearly seen that the MXP plot for node $1$ is non-monotonic. Moreover, MXP is unable to capture the global maximum value of the network (as some of the nodes have state values greater than MXP between iterations $0$ and $3$).

\section{Conclusions}\label{sec6}
In this article, we present a protocol for distributed finite-time termination of consensus algorithms in networks with dynamic topology. 
We introduced a novel concept of time-path which helps in analyzing the influence of an agent in the network on all other agents. We prove the existence of finite-length time-paths and establish the strict monotonic property of the global maximum and minimum of the ratio of the two state values of each node after every $n'$ (upper bound on number of nodes) number of iterations. 
A new Maximum-Minimum protocol for dynamic topology is presented and utilized to design the distributed finite-time termination algorithm. 
We discussed that the existing algorithms for finite-time termination based on static topology can fall short in case of dynamic topology for real-time applications such as \textit{ad hoc} cognitive radio networks and control of autonomous agents.
The effectiveness of our algorithm in these cases is demonstrated by experimentally realizing a network with dynamic topology created using a NodeJS framework. 

\bibliographystyle{ieeetr} 

\bibliography{topident}

\end{document}